\documentclass[aps,prd,amssymb,twocolumn,nofootinbib,10pt]{revtex4-1}
\usepackage{graphicx}
\usepackage{dcolumn}%
\usepackage{dsfont}
\usepackage{relsize}
\usepackage{scalerel}
\usepackage{bm}%
\usepackage{amssymb}
\usepackage[]{complexity}
\usepackage[version=3]{mhchem}
\usepackage{slashed}

\usepackage[bookmarks=false]{hyperref}
\newcommand{\RN}[1]{%
  \textup{\uppercase\expandafter{\romannumeral#1}}%
}
\begin{document}


\title{Target-normal single-spin asymmetry in elastic electron-nucleon scattering}


\author{Oleksandr Koshchii}
\email[]{koshchii@gwmail.gwu.edu}
\affiliation{The George Washington University, Washington, D.C. 20052, USA}

\author{Andrei Afanasev}

\affiliation{The George Washington University, Washington, D.C. 20052, USA}

\date{\today}

\begin{abstract}
We estimate the target-normal single-spin asymmetry at near forward angles in elastic electron-nucleon scattering. In the leading-order approximation, this asymmetry is proportional to the imaginary part of the two-photon exchange (TPE) amplitude, which can be expressed as an integral over the doubly virtual Compton scattering (VVCS) tensor. We develop a model that parametrizes the VVCS tensor for the case of near forward scattering angles. Our parametrization ensures a proper normalization of the imaginary part of the TPE amplitude on the well-known forward limit expression, which is given in terms of nucleon structure functions measurable in inelastic electron-nucleon scattering experiments. We discuss applicability limits of our theory and provide target-normal single-spin asymmetry predictions for both elastic electron-proton and electron-neutron scattering.
\end{abstract}

\maketitle

\section{Introduction}
Elastic lepton scattering off of a nucleon ($l^\pm N \rightarrow l^\pm N$) provides a great deal of information on the structure of the hadron. High precision and increasing accuracy of modern lepton scattering measurements push theoretical calculations beyond the leading order Born approximation. As a result, since the beginning of this century, many efforts have been devoted to improving our understanding of higher order contributions to elastic lepton-nucleon scattering, and two-photon exchange (TPE) corrections in particular. Hereafter, we just briefly discuss theoretical and experimental progress in understanding of the two-photon physics, whereas detailed reviews can be found in Refs. \cite{CarlsonTPE2007, ArringtonTPE2011, AfanasevTPE2017}.

Most of recent attempts to reexamine older treatments of radiative corrections in unpolarized electron-proton scattering have been triggered by the so-called ``proton form factor puzzle.'' This puzzle constitutes the discrepancy between the proton electric-to-magnetic form factor ratio $G_E^p / G_M^p$ measured in unpolarized and polarized \cite{JonesFF2000, GayouFF2002} electron-proton scattering at momentum transfers $Q^2 \gtrsim 1$ GeV$^2$. As it is suggested in Refs. \cite{BlundenTPE2003, GuichonTPE2003, ArringtonTPE2007}, the discrepancy can largely be mitigated if one accounts for hard TPE processes\footnote{The separation of a photon's phase space into the ``soft'' and ``hard'' regions is ambiguous. The most common conventions are those of Tsai \cite{TsaiRadCor1961} and Maximon and Tjon \cite{MaximonRadCor2000}.} in unpolarized measurements. However, corresponding  theoretical computations are dependent on the structure of the proton and have kinematical limitations. For instance, the hadronic (direct loop) calculations \cite{BlundenTPE2003, BlundenTPE2005} feature an undesired divergence in the high energy limit, whereas the partonic estimations \cite{ChenTPE2004, AfanasevTPE2005} are limited to the kinematical region where $Q^2 \gtrsim 1$ GeV$^2$. For the present, there exists no complete calculation of hard TPE that is valid at all kinematics.

Another problem that have furthered interest in the physics of TPE was the so-called ``proton radius puzzle'' \cite{PohlRadPuzzle2010, AntogniniRadPuzzle2013}. This puzzle encapsulates the difference between the radius of the proton as measured with electron scattering and atomic hydrogen spectroscopy, and that measured in muonic hydrogen spectroscopy. Regardless of the technique implemented, one needs to have a good understanding of TPE mechanisms in order to precisely determine the radius; for the detailed discussion on the extraction of the corresponding quantity from unpolarized electron-proton scattering and atomic spectroscopy, see Refs. \cite{HillRadius2010, BernauerRadius2010, BernauerRadius2014} and Refs. \cite{HillRadius2011, CarlsonProtRad2015}, respectively.

Not only did the two puzzles stimulate theoretical progress they also have given rise to multiple precision measurements of TPE. For example, recent VEPP-3, OLYMPUS, and CLAS experiments \cite{RachekTPE2015, HendersonTPEOlympus2017, RimalTPE2017} studied the real (dispersive) part\footnote{Note that, depending on an experimental design of a certain elastic lepton-proton scattering experiment, one may access only a real or an imaginary part of the TPE amplitude.} of the TPE amplitude, while they were looking for direct evidence of hard TPE. As it is well known, the corresponding contribution can be directly extracted from the ratio $R =d \sigma(l^+ p) / d \sigma(l^- p)$ of unpolarized scattering cross sections. The experiments \cite{RachekTPE2015,  HendersonTPEOlympus2017, RimalTPE2017} employed $e^\pm p$ scattering for their analysis of TPE and covered a wide kinematical range of $Q^2$ ($0.165 < Q^2 < 2.038$ GeV${}^2$). In addition to these measurements, the forthcoming MUSE experiment \cite{GilmanMUSE2013}, which is designed to be sensitive to the proton's radius, is going to study TPE in the low-$Q^2$ region ($0.0016 < Q^2 < 0.082$ GeV${}^2$). The relevance of this measurement will be assured by the respective experimental setup that enables a first simultaneous determination of TPE from unpolarized $e^\pm p$ and $\mu^\pm p$ scattering. Moreover, the kinematics of MUSE will provide means for precision studies of lepton mass effects in elastic lepton-proton scattering. As a result, the experiment has the potential to demonstrate whether the muon-proton and electron-proton interactions are different, and will check whether any differences are coming from novel physics or hard TPE; corresponding theoretical analysis that enables a proper extraction of hard TPE from elastic scattering of massive leptons off a proton target can be found in Ref. \cite{KoshchiiAsymmetry2017}.

Unlike the dispersive part of the TPE amplitude, the corresponding imaginary (absorptive) part\footnote{Note that the imaginary part of the TPE amplitude is contained solely in the two-photon box diagram, whereas the respective real part is represented by the two-photon box and crossed-box diagrams.} manifests itself in polarized scattering measurements. More specifically, it can be directly accessed through the analysis of a single-spin asymmetry (SSA) observable in elastic lepton-nucleon scattering, when either the beam or target is polarized in the direction normal to the lepton scattering plane. A respective theoretical investigation was performed several decades ago by De Rujula \textit{et al.} in Ref. \cite{RujulaSSA1971}. In that paper the authors explain why the transverse SSA must be zero in the Born approximation by considering electron scattering on a polarized proton target. Moreover, they have shown that the leading-order contribution to such an asymmetry is generated by the absorptive part of the TPE amplitude, which, in its turn, has drawn a significant theoretical interest in recent years. This interest is assured by the rapid development of dispersive methods in calculations of TPE \cite{GorchteinDispersive2007, BorisyukDisp2008, TomalakDisp2015, BlundenDisp2017}. Alternatively to the hadronic or partonic approaches, which suggest a direct calculation of the real part of the TPE contribution, the dispersive technique prescribes the evaluation of the imaginary contribution in the first place. A respective calculation should be performed by utilizing the unitarity property of the scattering matrix. As a result, one gets an exclusive opportunity to employ the on-shell form factor parametrization in their calculations of TPE. Once the imaginary part is computed, the corresponding real part can be reconstructed by making use of dispersive relations. As a consequence, the dispersive treatment allows for a meaningful reduction of theoretical uncertainties in calculations of the real part of the TPE contribution. It is also worth mentioning here that implications of TPE are important for a precision extraction of the proton's weak charge from parity-violating electron-proton scattering \cite{AfanasevGammaZ2005, GorchteinGammaZ2011, CarlsonGammaZ2011, HallGammaZ2016}.

Experimental capabilities to measure nonzero transverse SSAs were achieved relatively recently - about 15 years ago. Here we should note that a target-normal SSA ($A_y^N$) in elastic $e^\pm N$ scattering is usually expected to be of order $\alpha \approx 1/137$ (more details can be found in Sec. \ref{1.500}), whereas a beam-normal asymmetry ($B_y^N$) is expected to be about a thousand times smaller due to its additional proportionality to the beam's mass-to-energy ratio. Despite being relatively small, the beam-normal asymmetry was the first transverse SSA observed experimentally \cite{SampleBNSSA2001}. Subsequent measurements \cite{A4Mass2005, G0BNSSA2007, A4Capozza2007, G0BNSSA2011} also studied $B_y^p$, but in different kinematical settings. Moreover, the HAPPEX experiment at Jefferson Lab \cite{HAPPEX2012} not only accessed $B_y^p$, but also $B_y^{\ce{^4_{}He}}, B_y^{\ce{^{12}_{}C}},$ and $B_y^{\ce{^{208}_{}Pb}}$. As for the target-normal SSA, there is only one nonzero measurement of $A_y$ reported to date \cite{ZhangSSA2015}, which was obtained from quasielastic electron scattering on a polarized ${}^3$He nucleus. In addition to providing $A_y^{\ce{^3_{}He}}$, the authors of Ref. \cite{ZhangSSA2015} extract a nonzero neutron-normal SSA by using the effective neutron polarization approximation. The results of their measurement indicate that the neutron-normal SSA at GeV beam energies and $Q^2 \le 1$ GeV$^2$ is dominated by the inelastic TPE loop contribution (when the intermediate hadronic state is not given by the neutron).

As noted previously, early theoretical calculations of SSAs were performed for the case of the transversely polarized proton target \cite{RujulaSSA1971, RujulaSSA1973}. The authors of Ref. \cite{AfanasevTNSSA2002} improved on those near forward angle calculations of $A_y^p$ by accounting for the proton structure effects. Additionally, the formalism to describe $A_y^N$ at large momentum transfers ($Q^2 \gtrsim 1$ GeV$^2$) was developed in Ref. \cite{ChenTPE2004}. Similarly, the approach to address $B_y^N$ at large momentum transfers was provided in Ref. \cite{GorsteinBNSSA2004}. Moreover, the analytical behavior of $B_y^N$ in scattering at near forward angles was studied in Ref. \cite{AfanasevBNSSA2004}, and beyond forward angles in Ref. \cite{BorisyukBNSSA2006}. The Regge region behavior and analizing power of $B_y$ were considered in Refs. \cite{GorsteinBNSSA2006} and \cite{GorsteinBNSSA2008}, respectively. Pasquini and Vanderhaeghen analyzed the full angular behavior of $A_y^N$ and $B_y^N$ by making use of a phenomenological model that employs $\gamma^\star N \rightarrow \pi N$ electroproduction amplitudes \cite{PasquiniSSA2004}.

To our best knowledge, the only models that include the inelastic TPE loop contribution and have been used to predict the neutron-normal SSA are those of Refs. \cite{ChenTPE2004} and \cite{PasquiniSSA2004}. The generalized parton distributions (GPDs) calculation of the former reference agrees well with the experimental datapoint of Ref. \cite{ZhangSSA2015}, which was taken at electron beam energy $\varepsilon_1 = 3.605$ GeV and $Q^2 = 0.967$ GeV$^2$. However, this approach cannot be used to describe other datapoints of Ref. \cite{ZhangSSA2015}, which were taken at lower $Q^2$. The electroproduction amplitudes calculation of Ref. \cite{PasquiniSSA2004}, in its turn, is constrained by the invariant mass of the intermediate hadronic state $W \lesssim 2$ GeV (corresponds to $\varepsilon_1 \lesssim 1.66$ GeV). The neutron-normal SSA prediction of Ref. \cite{PasquiniSSA2004} is given only for $\varepsilon_1 = 0.57$ GeV. It is the goal of our work to provide additional model estimations of neutron- and proton-normal SSAs that can cover kinematical regions unaccessible by the mentioned models. Broadly speaking, our theory is aimed to describe near forward scattering angle asymmetries. To be more specific than just vaguely mentioning near forward angles as a kinematical constraint on our approach, we will formulate a quantitative criterium that can be used to determine applicability limits of our theory.

The outline of this work is as follows. In Sec. \ref{1.100}, we introduce our notations for the description of elastic electron-nucleon scattering. In Sec. \ref{1.200}, we show that the respective target-normal SSA is generated (to leading order) by the imaginary part of the TPE amplitude. In Sec. \ref{1.300}, we give a brief overview of one- and two-photon exchange contributions needed for calculations of nucleon-normal SSAs. In Sec. \ref{1.500}, we provide our parametrization and closed form expressions for calculations of corresponding asymmetries. The results and conclusions are presented in Secs. \ref{1.900} and \ref{1.1000}, respectively.

\section{Elastic electron-nucleon scattering formalism}\label{1.100}

\begin{figure}[htp]
    \includegraphics[scale=0.4]{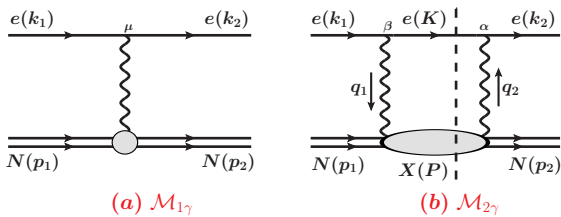}
    \caption{\label{fig:1}One- and two-photon exchange diagrams for elastic electron-nucleon scattering.}
\end{figure}

In this section we briefly review our notations that we use to describe the elastic electron-nucleon scattering. Schematically, this process, which is depicted in Fig. \ref{fig:1}, can be written as
\begin{equation}\label{1.101}
    e(k_1, S_e) + N(p_1, S_N) \rightarrow e(k_2, S'_e) + N(p_2, S'_N),
\end{equation}
where $k_1 (k_2)$ and $p_1 (p_2)$ denote the four-momenta of the initial (final) electron of mass $m$ and initial (final) nucleon of mass $M$. In addition, $S_e$ $(S'_e)$ and $S_N$ $(S'_N)$ describe the respective initial (final) electron and nucleon spin four-vectors. In order to provide invariant expressions, the standard set of Mandelstam variables is used
\begin{equation}\label{1.102}
    s = (k_1 + p_1)^2, \ t = (k_1 - k_2)^2, u = (k_1 - p_2)^2.
\end{equation}

Often, we shall refer to the absorptive part of the two-photon exchange amplitude, the definition for which is provided in Sec. \ref{1.200}, and the respective TPE Feynman diagram for which is shown in Fig. \ref{fig:1}(b). As we can see from this figure, the four-momentum of the intermediate electron state is denoted as $K$, so that $K^2 = m^2$, and the total energy-momentum of the intermediate hadronic state $X$ is denoted as $P$, so that the invariant mass squared $W^2$ of this state is then given by $P^2 = W^2$. The four-momenta of the virtual photons in Fig. \ref{fig:1} are given as
\begin{equation}\label{1.103}
\begin{split}
    q_1^2 & = (k_1 - K)^2 = (P - p_1)^2 \equiv - Q_1^2, \\
    q_2^2 & = (k_2 - K)^2 = (P - p_2)^2 \equiv - Q_2^2, \\
    t & = (q_1 - q_2)^2 \equiv - Q^2.
\end{split}
\end{equation}
Furthermore, it is convenient to introduce the following variables:
\begin{equation}\label{1.104}
    \bar{p} \equiv \frac{p_1 + p_2}{2}, \ \bar{q} \equiv \frac{q_1 + q_2}{2}.
\end{equation}

\section{Target-normal single-spin asymmetry}\label{1.200}

\begin{figure}[htp]
    \includegraphics[scale=0.42]{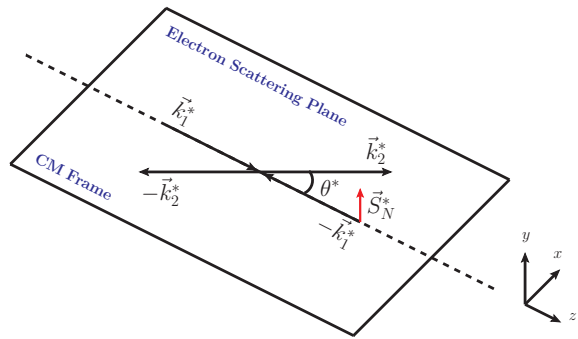}
    \caption{\label{fig:2} The center-of-mass frame used to describe the target-normal single-spin asymmetry in elastic $e N$ scattering.}
\end{figure}

In this section, in order to address the scattering process (\ref{1.101}), we refer to the center-of-mass (c.m.) frame, oriented as it is shown in Fig. \ref{fig:2}. In our notations, all c.m. frame variables always bear an asterisk symbol and correspond to analogous laboratory frame variables that do not bear this symbol. The complete list of our c.m. frame conventions and some useful invariant expressions are given in Appendix \ref{1.600}.

The target-normal single-spin asymmetry observable $A_y^N$ in elastic $e N$ scattering is defined as
\begin{equation}\label{1.201}
\begin{split}
    A_y^N & \equiv \frac{d \sigma_{N^{{}^\uparrow}} - d \sigma_{N^{{}^\downarrow}}}{d \sigma_{N^{{}^\uparrow}} + d \sigma_{N^{{}^\downarrow}}},
\end{split}
\end{equation}
where $d \sigma_{N^{{}^\uparrow}}$ ($d \sigma_{N^{{}^\downarrow}}$) denotes the differential cross section for the unpolarized electron beam and for the polarized target nucleon with spin vector $\vec{S}_N^*$ being oriented parallel (antiparallel) to the normal ($\vec{e}^*_y$) to the electron scattering plane and being normalized to 1. The four-vector spin is then given by
\begin{equation}\label{1.202}
    S^\mu_{N} = (0, \vec{S}_N^*), \ \ \ \ \vec{S}_N^* \equiv \frac{\vec{k}_1^* \times \vec{k}_2^*}{|\vec{k}_1^* \times \vec{k}_2^*|} = \vec{e}^*_y.
\end{equation}
In the tensor notation
\begin{equation}\label{1.203}
    S^\mu_{N} = \frac{1}{N_s} \varepsilon^{\mu \nu \rho \sigma} p_{1\nu} k_{1 \rho} k_{2\sigma},
\end{equation}
where the normalization constant $N_s$ is introduced to satisfy the condition $S^2_N = -1$. For scattering of ultrarelativistic  ($m \rightarrow 0$) electrons
\begin{equation}\label{1.204}
\begin{split}
    N_s & = \frac{1}{2} \sqrt{Q^2 \big[(M^2 - s)^2 - s Q^2 \big]}.
\end{split}
\end{equation}

Let us now define $T_{fi} \equiv T_{N^{{}^\uparrow}} (\vec{k}_2^*, \vec{k}_1^*)$ to be the transition amplitude describing the scattering process shown in Fig. \ref{fig:2}, so that
\begin{equation}\label{1.205}
    d \sigma_{N^{{}^\uparrow}} \sim \frac{1}{2} \sum \limits_{\scaleto{S'_N, S'_e, S_e}{5pt}}^{} |T_{fi}|^2.
\end{equation}
In addition, we define $T_{\tilde{f} \tilde{i}}$ to be the amplitude describing the analogous process, but reversed in time (the nucleon's spin vector and particles' momenta are flipped). As it was pointed out by de Rujula \textit{et al.} in Ref. \cite{RujulaSSA1971}
\begin{equation}\label{1.206}
\begin{split}
    |T_{\tilde{f} \tilde{i}}|^2 & \equiv \big|T_{N^{{}^\downarrow}} (- \vec{k}_2^*, - \vec{k}_1^*)\big|^2 = \big|e^{i \pi} \cdot T_{N^{{}^\downarrow}} (\vec{k}_2^*, \vec{k}_1^*)\big|^2 \\
    & = \big|T_{N^{{}^\downarrow}} (\vec{k}_2^*, \vec{k}_1^*)\big|^2.
\end{split}
\end{equation}
Using our definition of $d \sigma_{N^{{}^\downarrow}}$ and the result of Eq. (\ref{1.206}), one can find that
\begin{equation}\label{1.207}
    d \sigma_{N^{{}^\downarrow}} \sim \frac{1}{2} \sum \limits_{\scaleto{S'_N, S'_e, S_e}{5pt}}^{} |T_{\tilde{f} \tilde{i}}|^2.
\end{equation}
This means that the asymmetry $A_y^N$ can now be written as
\begin{equation}\label{1.208}
\begin{split}
    A_y^N & = \frac{|T_{fi}|^2 - |T_{\tilde{f} \tilde{i}}|^2}{|T_{fi}|^2 + |T_{\tilde{f} \tilde{i}}|^2},
\end{split}
\end{equation}
where the summation over respective spin states is assumed in the numerator and denominator. To study Eq. (\ref{1.208}) in more detail, let us now write down the relation between the scattering matrix $S_{fi}$ and the amplitude $T_{fi}$
\begin{equation}\label{1.209}
    S_{fi} = \mathds{1} + i (2 \pi)^4 \delta^{(4)} \big( k_1 + p_1 - k_2 - p_2 \big) T_{fi}.
\end{equation}
The unitarity property $S_{fi}^\dag S_{fi} = \mathds{1}$ of the scattering matrix enables us to find that
\begin{equation}\label{1.210}
    i \Big( T^\dag_{fi} - T_{fi} \Big) = \mathrm{Abs} \big [ T_{fi} \big],
\end{equation}
where $\mathrm{Abs} \big [ T_{fi} \big]$ is the absorptive part of the scattering amplitude, defined as
\begin{equation}\label{1.211}
    \mathrm{Abs} \big [ T_{fi} \big] \equiv \sum_{n}^{} T^*_{f n} T_{n i} (2 \pi)^4 \delta^{(4)}\big( k_1 + p_1 - p_n \big).
\end{equation}
The sum in Eq. (\ref{1.211}) goes over all possible on-shell intermediate states $n$, and the delta function there assures conservation of momentum. Using Eq. (\ref{1.210}), one may find that
\begin{equation}\label{1.212}
    \big| \mathrm{Abs} \big [ T_{fi} \big] \big|^2 = |T_{\tilde{f} \tilde{i}}|^2 + |T_{fi}|^2 - 2 \ \mathrm{Re} [T_{fi} T_{fi}], \\
\end{equation}
\begin{equation}\label{1.213}
    2 \ \mathrm{Im} \Big( T^\dag_{fi} \mathrm{Abs} \big [ T_{fi} \big] \Big) =  2 \ \mathrm{Re} [T_{fi} T_{fi}] - 2 |T_{\tilde{f} \tilde{i}}|^2,
\end{equation}
where, due to time-reversal invariance and parity conservation in the electromagnetic interaction, we replaced $|T_{if}|^2$ by $|T_{\tilde{f} \tilde{i}}|^2$. Now, taking into account that
\begin{equation}\label{1.214}
\begin{split}
    T_{fi} & = \big( T_{1 \gamma} \big)_{fi} + \big( T_{2 \gamma} \big)_{fi} + ...
\end{split}
\end{equation}
and summing up Eqs. (\ref{1.212}) and (\ref{1.213}), one gets a perturbative expansion of the numerator in Eq. (\ref{1.208}). As a result, the dominant contribution (of order $\alpha$) to the asymmetry will be given by the following expression:
\begin{equation}\label{1.215}
    A_y^N = \frac{\mathrm{Im} \Big[ \big( T_{1 \gamma} \big)^\dag_{fi} \cdot \big( \mathrm{Abs} \big [ T_{2 \gamma} \big] \big)_{fi} \Big]}{ \big| \big( T_{1 \gamma} \big)_{fi} \big|^2},
\end{equation}
where we employed the equivalence of $ \big( T_{1 \gamma} \big)_{\tilde{f} \tilde{i}}$ and $\big( T_{1 \gamma} \big)_{fi}$ in the denominator of Eq. (\ref{1.208}). As it is mentioned in Ref. \cite{Akhiezer1977book}, in the one-photon exchange approximation, the differential cross section of an unpolarized lepton scattering off of a polarized target is identical to that of analogous process but with the target being unpolarized. With this in mind, and dropping out the matrix indices in Eq.(\ref{1.215}), one gets \cite{PasquiniSSA2004}
\begin{equation}\label{1.216}
    A_y^N = \frac{2 \ \mathrm{Im} \bigg( \sum \limits_{\scaleto{S'_N, S'_e, S_e}{5pt}}^{} T_{1 \gamma}^\dag \cdot \mathrm{Abs} \big [ T_{2 \gamma} \big] \bigg)} {\sum \limits_{\scaleto{S'_N, S'_e, S_N, S_e}{5pt}}^{} \big| \tilde{T}_{1 \gamma} \big|^2},
\end{equation}
where $\tilde{T}_{1 \gamma}$ is the one-photon exchange amplitude describing the unpolarized scattering process, characterized by the differential cross section $d \sigma_{un}$, which is given by
\begin{equation}\label{1.217}
    d \sigma_{un} \sim \frac{1}{4} \sum \limits_{\scaleto{S'_N, S'_e, S_N, S_e}{5pt}}^{} |\tilde{T}_{fi}|^2.
\end{equation}
At this point, it is worth mentioning that the result of Eq. (\ref{1.216}) holds true if one works in the laboratory (lab) frame where the initial nucleon is motionless and the initial electron moves along the $z$ axis. This stems from the fact that such a lab frame can be obtained from the c.m. frame by the inverse Lorentz boost in the $z$ direction, thus keeping the components of the spin four-vector unchanged.

\section{One- and two-photon exchange contributions}\label{1.300}

The one-photon exchange amplitude, which is shown in Fig. \ref{fig:1}(a) and contributes to the asymmetry in Eq. (\ref{1.216}), is given by
\begin{equation}\label{1.301}
    T_{1 \gamma}^\dag = \frac {e^2}{Q^2} \bar{u}(k_1, S_e) \gamma^\mu u(k_2, S'_e) \bar{U}(p_1, S_N) \Gamma_\mu U(p_2, S'_N).
\end{equation}
The on-shell nucleon vertex $\Gamma_\mu$ is defined as
\begin{equation}\label{1.302}
    \Gamma_\mu (Q^2) = \Big[F_1 (Q^2) + F_2 (Q^2) \Big] \gamma_\mu - \frac{\bar{p}_\mu}{M} F_2 (Q^2),
\end{equation}
where $F_1$ and $F_2$ are the Dirac and Pauli form factors, which are related to the electric $G_E$ and magnetic $G_M$ Sachs form factors via
\begin{equation}\label{1.303}
\begin{split}
    G_E (Q^2) & = F_1(Q^2) - \frac{Q^2}{4 M^2} F_2(Q^2), \\
    G_M (Q^2) & = F_1(Q^2) + F_2(Q^2).
\end{split}
\end{equation}
In our calculations of proton- and neutron-normal SSAs we use Kelly's parametrisation \cite{KellyFF2004} to describe the respective $Q^2$ behavior of the electric and magnetic form factors. For the proton, the fit parameters were taken from Ref. \cite{KellyFF2004}, whereas neutron's $G_E$ and $G_M$ fit parameters were taken from Refs. \cite{WojtsekhowskiFF2010} and \cite{QattanFF2012}, correspondingly.

As mentioned previously, the denominator in Eq. (\ref{1.216}) represents the one-photon exchange amplitude of the unpolarized electron-nucleon scattering. The square of this amplitude, summed over final and averaged over initial spins, can be written as
\begin{equation}\label{1.304}
    \frac{1}{4} \sum \limits_{\scaleto{S'_N, S'_e, S_N, S_e}{5pt}}^{} \big| \tilde{T}_{1 \gamma} \big|^2 = \frac{64 \pi^2 \alpha^2}{Q^4} D(s, Q^2),
\end{equation}
with
\begin{equation}\label{1.305}
\begin{split}
    D(s, Q^2) & \equiv \Big( (s - M^2)^2 - s Q^2 \Big) \Big( F_1^2 + \frac{Q^2}{4 M^2} F_2^2 \Big) \\
              & + \frac{Q^4}{2} \Big( F_1 + F_2 \Big)^2.
\end{split}
\end{equation}

The absorptive part of the TPE amplitude is connected with a discontinuity of the Feynman diagram Fig. \ref{fig:1}(b) via \cite{Peskin:1995ev}
\begin{equation}\label{1.401}
     \mathrm{Abs} \big[ T_{2 \gamma} \big] = - \mathrm{Disc} \big[ i T_{2 \gamma} \big] \equiv - \mathrm{Disc} \big[ \mathcal{M}_{2 \gamma} \big].
\end{equation}
This discontinuity can be calculated using the Cutkosky cutting rule prescription, which suggests that one replaces each cut propagator by the corresponding delta function
\begin{equation}\label{1.402}
     \frac{1}{p_i^2 - m_i^2 + i \epsilon} \rightarrow - 2 \pi i \delta(p_i^2 - m_i^2),
\end{equation}
where $m_i$ is the mass of the particle with the intermediate momentum $p_i$.

By calculating the discontinuity of Fig. \ref{fig:1}(b), one gets the following c.m. expression for the absorptive part of the TPE amplitude:
\begin{equation}\label{1.403}
\begin{split}
     \mathrm{Abs} \big[ T_{2 \gamma} \big] = e^4 \iiint & \frac{d^3 \vec{K}^*}{(2 \pi)^3 2 \xi^*} \frac{ W_{\alpha \beta} (p_2, S'_N; p_1, S_N)}{Q^2_1 Q^2_2}\\
     & \times \bar{u}(k_2, S'_e) \gamma^\alpha (\slashed{K} + m) \gamma^\beta u(k_1, S_e),
\end{split}
\end{equation}
where $\xi^*$ is the c.m. energy and $\vec{K}^*$ is the c.m. momentum of the intermediate electron (respective invariant form expressions are provided in Appendix \ref{1.600}). In addition, the TPE hadronic tensor $W_{\alpha \beta} (p_2, S'_N; p_1, S_N)$  is defined as
\begin{equation}\label{1.404}
\begin{split}
     W_{\alpha \beta} & (p_2, S'_N; p_1, S_N) \equiv \\
     & \sum \limits_X <p_2, S'_N| J^{\dag}_\alpha (0)|X> <X| J_{\beta} (0)|p_1, S_N> \\
     & \ \ \ \ \cdot (2 \pi)^4 \delta^4 \big( p_1 + q_1 - P \big).
\end{split}
\end{equation}
The sum in Eq. (\ref{1.404}) goes over all possible on-shell intermediate hadronic states $X$.

To perform a summation over final hadron spin states in Eq. (\ref{1.216}), it is convenient to relate the TPE hadronic tensor $W_{\alpha \beta}$ to an operator $\hat{W}_{\alpha \beta}$ in spin space, defined as \cite{CarlsonSSA2017}
\begin{equation}\label{1.406}
\begin{split}
     W_{\alpha \beta} (p_2, S'_N; p_1, S_N) \equiv \bar{U}(p_2, S'_N) \hat{W}_{\alpha \beta} (p_2, p_1) {U}(p_1, S_N).
\end{split}
\end{equation}
The tensor $\hat{W}_{\alpha \beta}$ corresponds to the absorptive part of the doubly virtual Compton scattering (VVCS) tensor $T_{\alpha \beta}$, so that
\begin{equation}\label{1.405}
     \hat{W}_{\alpha \beta} = \mathrm{Abs} \big[{T}_{\alpha \beta} \big] = 2 \ \mathrm{Im} \big[{T}_{\alpha \beta} \big].
\end{equation}
The absorptive part of the TPE amplitude can now be rewritten as
\begin{equation}\label{1.407}
\begin{split}
     \mathrm{Abs} \big[ T_{2 \gamma} \big] = e^4 \iiint & \frac{d^3 \vec{K}^*}{(2 \pi)^3 2 \xi^*} \frac{ \bar{U}(p_2, S'_N) \hat{W}_{\alpha \beta} {U}(p_1, S_N)}{Q^2_1 Q^2_2}\\
     & \times \bar{u}(k_2, S'_e) \gamma^\alpha (\slashed{K} + m) \gamma^\beta u(k_1, S_e).
\end{split}
\end{equation}

\section{Target-normal single-spin asymmetry calculation}\label{1.500}

By using the results of Eqs. (\ref{1.216}), (\ref{1.301}), (\ref{1.304}), and (\ref{1.407}), one may get the following expression for the target-normal SSA:
\begin{equation}\label{1.501}
    A_y^N (s, Q^2)= \frac{\alpha Q^2}{8 \pi^2 D(s, Q^2)} \iiint \frac{d^3 \vec{K}^*}{2 \xi^*} \frac{\mathrm{Im} \Big( L^{\mu \alpha \beta} H_{\mu \alpha \beta} \Big)}{Q_1^2 Q_2^2},
\end{equation}
where the leptonic $L^{\mu \alpha \beta}$ and hadronic $H_{\mu \alpha \beta}$ tensors are defined as
\begin{multline}\label{1.502}
    L^{\mu \alpha \beta} \equiv \frac{1}{2} \sum \limits_{S_e, S'_e} \bar{u}(k_1, S_e) \gamma^\mu u(k_2, S'_e) \bar{u}(k_2, S'_e) \gamma^\alpha \\
    \times (\slashed{K} + m) \gamma^\beta  u(k_1, S_e) \\
    = \frac{1}{2} \mathrm{Tr} \Big[ (\slashed{k}_1 + m) \gamma^\mu (\slashed{k}_2 + m) \gamma^\alpha (\slashed{K} + m) \gamma^\beta \Big],
\end{multline}
\begin{multline}\label{1.503}
    H_{\mu \alpha \beta} \equiv \sum \limits_{S'_N} \bar{U}(p_1, S_N) \Gamma_\mu U(p_2, S'_N) \\
    \times \bar{U}(p_2, S'_N) \hat{W}_{\alpha \beta} U(p_1, S_N) \\
    = \frac{1}{2} \mathrm{Tr} \Big[ (\slashed{p}_1 + M) (1 - \gamma_5 \slashed{S}_N) \Gamma_\mu (\slashed{p}_2 + M) \hat{W}_{\alpha \beta} \Big].
\end{multline}
It is useful to split the leptonic tensor Eq. (\ref{1.502}) into the symmetric ($L^{\mu \alpha \beta}_S$) and antisymmetric ($L^{\mu \alpha \beta}_A$) parts with respect to indices $\alpha \beta$. For ultrarelativistic electrons these parts are given by
\begin{equation}\label{1.550}
\begin{split}
    L^{\mu \alpha \beta}_S & = 2 {k}_2^{\mu } \left({K}^{\alpha } {k}_1^{\beta }+{k}_1^{\alpha } {K}^{\beta }\right)+2 {k}_1^{\mu } \left({K}^{\alpha } {k}_2^{\beta }+{k}_2^{\alpha } {K}^{\beta }\right)\\
    & + g^{\alpha  \beta } \Big(Q^2 {K}^{\mu } + q_1^2 {k}_2^{\mu } + q_2^2 {k}_1^{\mu } \Big) \\
    & - Q^2 \Big({K}^{\alpha } {g}^{\beta  \mu }+{K}^{\beta } {g}^{\alpha  \mu } \Big),
\end{split}
\end{equation}
\begin{equation}\label{1.551}
\begin{split}
    L^{\mu \alpha \beta}_A & = q_1^2 \left( {k}_2^{\alpha} {g}^{\beta \mu} - {k}_2^{\beta} {g}^{\alpha \mu} \right) - q_2^2 \left({k}_1^{\alpha} {g}^{\beta \mu} - {k}_1^{\beta } {g}^{\alpha \mu} \right) \\
    & -2 {K}^{\mu} \left({k}_1^{\alpha} {k}_2^{\beta} - {k}_2^{\alpha} {k}_1^{\beta} \right).
\end{split}
\end{equation}

The integral over intermediate electron's phase space variables in Eq. (\ref{1.501}) can be reexpressed in a Lorentz invariant way through the following change of integration variables:
\begin{equation}\label{1.504}
\begin{split}
    \iint & d \Omega_{K^*} = 2 \int \limits_{-1}^{1} d \cos {\theta_1^*} \int \limits_{0}^{\pi} d \phi_1^* \\
    & = \frac{1}{\varepsilon^* \xi^*} \int \limits_{0}^{4 \varepsilon^* \xi^*} d Q_1^2 \int \limits_{Q_{-}}^{Q_{+}} \frac{d Q^2_2}{\sqrt{(Q_{+} - Q_2^2)(Q_2^2 - Q_{-})}}, \\
    \int \limits_{0}^{|\vec{K}_{max}^*|} & d |\vec{K}^*| = - \int \limits_{M^2}^{s} \frac{d W^2}{2 \sqrt{s}},
\end{split}
\end{equation}
where \cite{GorsteinBNSSA2008}
\begin{equation}\label{1.505}
\begin{split}
    Q_{\pm} & = \frac{\xi^*}{\varepsilon^*} Q^2 + Q_1^2 - \frac{Q^2 Q_1^2}{2 (\varepsilon^*)^2} \\
    & \pm 2 \sqrt{Q^2 Q_1^2} \sqrt{\frac{\xi^*}{\varepsilon^*} \Big( 1 - \frac{Q^2}{4 (\varepsilon^*)^2} \Big) \Big( 1 - \frac{Q_1^2}{4 \varepsilon^* \xi^*} \Big)}, \\
    |\vec{K}_{max}^*| & = \frac{s - M^2}{2 \sqrt{s}}. \\
\end{split}
\end{equation}
Here we should also mention that it is convenient to split the integral over the variable $W^2$ into two pieces
\begin{equation}\label{1.506}
    \int \limits_{M^2}^{s} (...) \ d W^2 = \int \limits_{M^2}^{(M + m_\pi)^2} (...) \ d W^2 + \int \limits_{(M + m_\pi)^2}^{s} (...) \ d W^2,
\end{equation}
where $m_\pi$ denotes the mass of a pion. The first integral on the right-hand side of Eq. (\ref{1.506}) describes the contribution that is coming from the so-called elastic intermediate hadronic state ($X = $ nucleon in the blob in Fig. \ref{fig:1}(b)), and we denote the tensor $\hat{W}_{\alpha \beta}$ under this integral as $\hat{W}_{\alpha \beta}^{el}$. The second integral on the right-hand side of Eq. (\ref{1.506}) describes the contribution that is coming from the so-called inelastic intermediate hadronic state ($X \neq $ nucleon in the blob in Fig. \ref{fig:1}(b)), and we denote the tensor $\hat{W}_{\alpha \beta}$ under this integral as $\hat{W}_{\alpha \beta}^{in}$. Once $\hat{W}_{\alpha \beta}^{el}$ and $\hat{W}_{\alpha \beta}^{in}$ are parametrized, the asymmetry Eq. (\ref{1.501}) can be calculated numerically using relations in Eqs. (\ref{1.502})-(\ref{1.506}). The details about our parametrizations for $\hat{W}_{\alpha \beta}^{el}$ and $\hat{W}_{\alpha \beta}^{in}$ are given below.

\subsection{Elastic Contribution}

By putting the intermediate nucleon on shell and using our definitions of ${W}_{\alpha \beta}$ and $\hat{W}_{\alpha \beta}$, one can explicitly express $\hat{W}_{\alpha \beta}^{el}$ through electromagnetic form factors of the nucleon
\begin{equation}\label{1.507}
\begin{split}
    \hat{W}_{\alpha \beta}^{el} = 2 \pi \delta(W^2 - M^2) \Gamma_\alpha (Q_2^2) (\slashed{P} + M) \Gamma_\beta (Q_1^2),
\end{split}
\end{equation}
where
\begin{equation}\label{1.508}
\begin{split}
    \Gamma_\beta (Q_1^2) & = \Big[ F_1 (Q_1^2) + F_2 (Q_1^2) \Big] \gamma_\beta - \frac{(P + p_1)_\beta}{2 M} F_2 (Q_1^2), \\
    \Gamma_\alpha (Q_2^2) & = \Big[ F_1 (Q_2^2) + F_2 (Q_2^2) \Big] \gamma_\alpha - \frac{(P + p_2)_\alpha}{2 M} F_2 (Q_2^2).
\end{split}
\end{equation}

\subsection{Inelastic Contribution}

In order to parametrize the inelastic tensor $\hat{W}_{\alpha \beta}^{in}$, we will make use of Eq. (\ref{1.405}), which relates $\hat{W}_{\alpha \beta}$ to the imaginary part of the VVCS tensor $T_{\alpha \beta}$,
\begin{equation}\label{1.580}
     \hat{W}_{\alpha \beta}^{in} = 2 \ \mathrm{Im} \big[{T}_{\alpha \beta} \big].
\end{equation}
It turns out \cite{Tarrach1975} that in the most general case of a scattering on a polarized nucleon target, the VVCS tensor is given in terms of a sum of $N = 18$ structures, consisting of gauge invariant tensors $(\tau_i)_{\alpha \beta}$ that have no kinematical singularities and corresponding independent amplitudes $A_i$ ($i = 1, ... , N$)\footnote{A ``structure'' is a product between $A_i$ and respective $(\tau_i)_{\alpha \beta}$.}. The original basis of 18 tensors was suggested by Tarrach in Ref. \cite{Tarrach1975}. However, that basis is ``nonminimal'', and its nonminimality implies that there is a linear dependence between some elements of the basis in a kinematical region where $(q_1 \cdot q_2) = 0$ is possible. To avoid this constraint, one may refer to a different basis. In our calculations, we used an alternative basis of Ref. \cite{DrechselComptonTens1998}, which has no kinematical singularities whenever it is employed to model the inelastic TPE hadronic tensor. In this basis, the VVCS tensor is given by
\begin{equation}\label{1.509}
\begin{split}
     T_{\alpha \beta} & = \sum \limits_{i \in J} (\tau_i)_{\alpha \beta} \ A_i \big( q_1^2, q_2^2, W^2, Q^2 \big), \\
     J & = \{ 1 , ... , 21 \} \backslash \{ 5, 15, 16 \}.
\end{split}
\end{equation}
Exact expressions for $(\tau_i)_{\alpha \beta}$ are given in Ref. \cite{DrechselComptonTens1998}, and they are based on the tensors of Tarrach. Unfortunately, in the most general case of a scattering at nonforward angles, the complete functional dependence of the amplitudes $A_i \big( q_1^2, q_2^2, W^2, Q^2 \big)$ is unknown. However, some of these amplitudes, especially those whose behavior in certain physical limits is wellunderstood, can be modeled realistically. For our calculations of elastic scattering at near forward angles ($Q^2 \ll s$) and for $Q^2 \lesssim 1$ GeV$^2$ we will need to employ the so-called forward scattering limit ($Q^2 = 0$ and $q_1^2 = q_2^2$). The relevant discussions about the configuration of $T_{\alpha \beta}$ in various limits can be found, e.g., in Refs. \cite{Drechsel2003, Hagelstein2016}. In the forward limit, the number of independent amplitudes $A_i$ and respective tensors $(\tau_i)_{\alpha \beta}$ has to reduce to $N = 4$. More specifically, there are two structures that characterize the spin-dependent part of $T_{\alpha \beta}$ and there are another two structures that describe the spin-independent part of ${T}_{\alpha \beta}$. The optical theorem allows to parametrize exactly imaginary parts of all forward amplitudes in terms of structure functions of the nucleon. These structure functions can be extracted from deep inelastic scattering (DIS) measurements and are introduced in Appendix \ref{1.700}. The exact expression for the forward VVCS tensor and the normalization of its imaginary part on the nucleon structure functions are provided in Appendix \ref{1.1100}.

Because of our limited knowledge about the functions $A_i$ from Eq. (\ref{1.509}), certain model assumptions need to be made. In the kinematical region of our interest (near forward scattering angles), we can assume that the leading role in the parametrization of ${T}_{\alpha \beta}$ is played by a sum of four structures that do not die off in the forward limit (FL); we call respective structures ``near forward'' contributions. The rest of the structures, which we call ``off-forward'' contributions, will be excluded from our model. Our assumption to consider only near forward contributions is based on the fact that amplitudes are smooth functions, meaning that the off-forward contributions can gain significance only continuously with the growth of $Q^2$ starting at $Q^2 = 0$, where they are irrelevant. Given this consideration, in our model for ${T}_{\alpha \beta}$ we intend to focus on the identification of linear combinations of four structures that contribute to the forward-limit parametrization of ${T}_{\alpha \beta}$. Two of these structures are spin dependent and the other two are spin independent. We can narrow down our searches for respective structures even further if we now analyze the behavior of the antisymmetric part of the leptonic tensor $L^{\mu \alpha \beta}_A$. Based on Eq. (\ref{1.551}), this part turns out to vanish in FL. This means that only the symmetric part $L^{\mu \alpha \beta}_S$ will be contributing to our model, as we just consider the structures that survive in the forward limit. Consequently, only the symmetric part $H_{\mu \alpha \beta}^S$ of the hadronic tensor needs to be employed in our parametrization,
\begin{equation}\label{1.552}
\begin{split}
     H_{\mu \alpha \beta}^S & = \frac{1}{2} \mathrm{Tr} \Big[ (\slashed{p}_1 + M) (1 - \gamma_5 \slashed{S}_N) \Gamma_\mu (\slashed{p}_2 + M) \hat{W}_{\alpha \beta}^{in, S} \Big] \\
     & \ = \mathrm{Tr} \Big[ (\slashed{p}_1 + M) (1 - \gamma_5 \slashed{S}_N) \Gamma_\mu (\slashed{p}_2 + M) \mathrm{Im}(\hat{T}_{\alpha \beta}) \Big],
\end{split}
\end{equation}
where we denoted the symmetric part of the VVCS tensor as $\hat{T}_{\alpha \beta}$. As it can be seen from the discussion in Appendix \ref{1.1100}, the symmetric part of the forward VVCS tensor is solely represented by the nucleon unpolarized structure functions. Consequently, our model will be focused on a near forward parametrization of ${T}_{\alpha \beta}$ in terms of \emph{two spin-independent} structures that contribute to forward scattering.\footnote{Even though we consider only the structures contributing to the forward limit, our model will still keep their dependence on $q_{1, 2}$ and $p_{1,2}$. Therefore, the wording ``near forward'' is chosen.}

Summing up the discussion above, from 18 tensors $(\tau_i)_{\alpha \beta}$ and functions $A_i \big( q_1^2, q_2^2, W^2, Q^2 \big)$ mentioned in Eq. (\ref{1.509}), we intend to identify those that can contribute to forward scattering and construct a sum (or sums) of two structures that would exactly reproduce Eqs. (\ref{1.1103}, \ref{1.1104}) in the forward limit kinematics. Essentially, from the set of 18 tensors provided by Ref. \cite{DrechselComptonTens1998}, we need to pin down those that can be inextricably linked to the forward spin-independent amplitudes. We found three tensors that reproduce the first term (its tensor part) of Eq. (\ref{1.1103}) and one tensor that reproduces the second term (its tensor part) of Eq. (\ref{1.1103}) in the forward limit. To be consistent with Ref. \cite{DrechselComptonTens1998}, these tensors are labeled as $(\tau_1)_{\alpha \beta}, (\tau_3)_{\alpha \beta}, (\tau_4)_{\alpha \beta}, (\tau_{19})_{\alpha \beta}$ and their exact form, using our notations, is provided in Appendix \ref{1.800}. Based on these tensors, we constructed three parametrizations
\begin{equation}\label{1.512}
\begin{split}
    \hat{T}_{\alpha \beta}^{(\RN{1})} & = (\tau_1)_{\alpha \beta} A_1 + (\tau_{19})_{\alpha \beta} A_{19}, \\
    \hat{T}_{\alpha \beta}^{(\RN{2})} & = (\tau_3)_{\alpha \beta} A_3 + (\tau_{19})_{\alpha \beta} A_{19}, \\
    \hat{T}_{\alpha \beta}^{(\RN{3})} & = (\tau_4)_{\alpha \beta} A_4 + (\tau_{19})_{\alpha \beta} A_{19}.
\end{split}
\end{equation}
In FL, which is characterized by the $q_1 = q_2 = q$ (or $p_1 = p_2 = p$) and $Q^2 = 0$ condition\footnote{Please note that $Q^2$ is the overall momentum transfer squared, whereas $q^2$ is the forward limit momentum square of a single photon in the TPE graph in Fig. \ref{fig:6}.}, the parametrizations $\hat{T}_{\alpha \beta}^{(\RN{1})}, \hat{T}_{\alpha \beta}^{(\RN{2})}, \hat{T}_{\alpha \beta}^{(\RN{3})}$ in Eq. (\ref{1.512}) exactly coincide with the spin-independent part of the forward VVCS amplitude Eq. (\ref{1.1101}) if the following constraints are imposed on the imaginary parts of the functions $A_1, A_3, A_4,$ and $A_{19}$:
\begin{equation}\label{1.513}
\begin{split}
     &\mathrm{Im} \big[ A_1 \big] \Big|_{\mathrm{FL}} = \mathrm{Im} \big[ A_1 \big( q^2, W^2 \big) \big] = \frac{\pi \ W_1^\mathrm{DIS} \big(q^2, W^2 \big)}{q^2}, \\
     &\mathrm{Im} \big[ A_3 \big] \Big|_{\mathrm{FL}} = \mathrm{Im} \big[ A_3 \big( q^2, W^2 \big) \big] = - \frac{\pi \ W_1^\mathrm{DIS} \big( q^2, W^2 \big)}{q^4}, \\
     &\mathrm{Im} \big[ A_4 \big] \Big|_{\mathrm{FL}} = \mathrm{Im} \big[ A_4 \big( q^2, W^2 \big) \big] = - \frac{\pi \ W_1^\mathrm{DIS} \big( q^2, W^2 \big)}{2 q^2 (p \cdot q)}, \\
     &\mathrm{Im} \big[ A_{19} \big] \Big|_{\mathrm{FL}} = \mathrm{Im} \big[ A_{19} \big( q^2, W^2 \big) \big] = \frac{\pi \  W_2^\mathrm{DIS} \big(q^2, W^2 \big)}{2 q^4 M^2}.
\end{split}
\end{equation}
$W_{1 (2)}^{\mathrm{DIS}}$ in Eq. (\ref{1.513}) represent the nucleon structure functions, which can be related to corresponding dimensionless scaling functions $F_{1 (2)}^{\mathrm{DIS}}$ measurable in deep inelastic scattering experiments (see Appendix \ref{1.700} for details),
\begin{equation}\label{1.515}
\begin{split}
     & \ W_{1}^{\mathrm{DIS}} \big( q_{1 (2)}^2, W^2 \big) = \frac{F_1^{\mathrm{DIS}} \big( q_{1 (2)}^2, x_{{B}_{1 (2)}} \big)}{M} , \\
     & \ W_{2}^{\mathrm{DIS}} \big( q_{1 (2)}^2, W^2 \big) = \frac{F_2^{\mathrm{DIS}} \big( q_{1 (2)}^2, x_{{B}_{1 (2)}} \big)}{\nu_{1 (2)}}, \\
     & \nu_{1 (2)} = \frac{(p_{1 (2)} \cdot q_{1 (2)})}{M} = \frac{W^2 - M^2 - q_{1 (2)}^2}{2 M}, \\
     & \ \ \ \ \ \ \ \ \ \ \ \ \ \ \ \ x_{{B}_{1 (2)}} = - \frac{q_{1 (2)}^2}{2 M \nu_{1 (2)}}.
\end{split}
\end{equation}
The constraints of Eq. (\ref{1.513}) are not the only requirements that should be taken into account in our model for the functions $A_1, A_3, A_4,$ and $A_{19}$. We also want to preserve the symmetry of these amplitudes under the exchange of momenta of virtual photons ($q_1^2 \leftrightarrow q_2^2$) in the TPE loop. Finally, a realistic $Q^2$ dependence should also be chosen to describe the behavior of the amplitude near FL. Based on these requirements, we suggest to parametrize the imaginary parts of near forward amplitudes as
\begin{equation}\label{1.5131}
\begin{split}
     &\mathrm{Im} \big[ A_1 (q_1^2, q_2^2, W^2, Q^2) \big] = \frac{\pi \ W_1 \big( q_1^2, q_2^2, W^2, Q^2 \big)}{(q_1 \cdot q_2)}, \\
     &\mathrm{Im} \big[ A_3 (q_1^2, q_2^2, W^2, Q^2) \big] = - \frac{\pi \ W_1 \big( q_1^2, q_2^2, W^2, Q^2 \big)}{q_1^2 q_2^2}, \\
     &\mathrm{Im} \big[ A_4 (q_1^2, q_2^2, W^2, Q^2) \big] = - \frac{\pi \ W_1 \big( q_1^2, q_2^2, W^2, Q^2 \big)}{(\bar{p} \cdot \bar{q}) (q_1^2 + q_2^2)}, \\
     &\mathrm{Im} \big[ A_{19} (q_1^2, q_2^2, W^2, Q^2) \big] = \frac{\pi \  W_2 \big( q_1^2, q_2^2, W^2, Q^2 \big)}{2 q_1^2 q_2^2 M^2},
\end{split}
\end{equation}
where, following the discussion in Ref. \cite{RujulaSSA1971}, the structures $W_1$ and $W_2$ are modeled as
\begin{equation}\label{1.514}
     W_{1 (2)} = e^{- \frac{B Q^2}{2}} \sqrt{W_{1 (2)}^{\mathrm{DIS}} \big( q_1^2, W^2 \big) W_{1 (2)}^{\mathrm{DIS}} \big( q_2^2, W^2 \big)}.
\end{equation}
The coefficients by the functions $W_{1(2)}$ in Eq. (\ref{1.5131}) are obtained based on the constraints of Eq. (\ref{1.513}). The model parametrization of the structures $W_{1 (2)}$ given by Eq. (\ref{1.514}) ensures symmetry considerations with respect to virtual photon exchanges in the TPE loop. The exponent in Eq. (\ref{1.514}) accounts for the nucleon-size effects and is introduced to describe the $Q^2$ behavior of the scattering amplitude near the forward scattering limit. The experimentally determined constant $B = 8$ GeV${}^{-2}$, which is obtained from the slope of the Compton scattering amplitude, gives a good description of near forward scattering up to $Q^2 \approx 1$ GeV${}^2$; for more details on the exponential behavior of the amplitude near FL, please see Refs. \cite{AfanasevBNSSA2004, GorsteinBNSSA2008}.

One may wonder why we prefer to employ the geometric mean over, e.g., the arithmetic mean to model $W_{1 (2)}$ in Eq. (\ref{1.514}). Besides the argument of Ref. \cite{RujulaSSA1971}, the choice of the geometric mean is driven by the analysis of the behavior of the nonforward Compton amplitude. Based on both the vector meson dominance model (for low $Q_{1(2)}^2$) and quark counting rules and experimental measurements (for high $Q_{1(2)}^2$), the nonforward Compton amplitude is supposed to decrease at the fixed value of the virtuality of one of the photons but increasing virtuality of the other photon in the TPE loop. Let us now assume that $Q_1^2$ is fixed and $Q_2^2$ is increasing. The arithmetic mean model, in this case, would not follow the desired trend assuming that the term proportional to $Q_1^2$ dominates over its counterpart proportional to $Q^2_2$. In this scenario, the change in $Q^2_2$ would not, essentially, lead to the change of the Compton amplitude, thus providing an overestimation of the integral over $Q_1^2$ and $Q_2^2$ (Eqs. (\ref{1.501}) and (\ref{1.504})). Consequently, we expect the SSA prediction with the arithmetic mean parametrization of structure functions to be less accurate than that with the geometric mean parametrization. As for the forward limit calculation, both parametrizations would be equivalent.

On a final note we would like to mention that in our calculations of nucleon-normal SSAs we used Christy's \cite{Christy} parametrizations of $F_{1}^{\mathrm{DIS}}$ and $F_{2}^{\mathrm{DIS}}$, which include the nucleon resonance region.

\section{Results and Discussion}\label{1.900}

\begin{figure*}[tp]
    \includegraphics[scale=0.45]{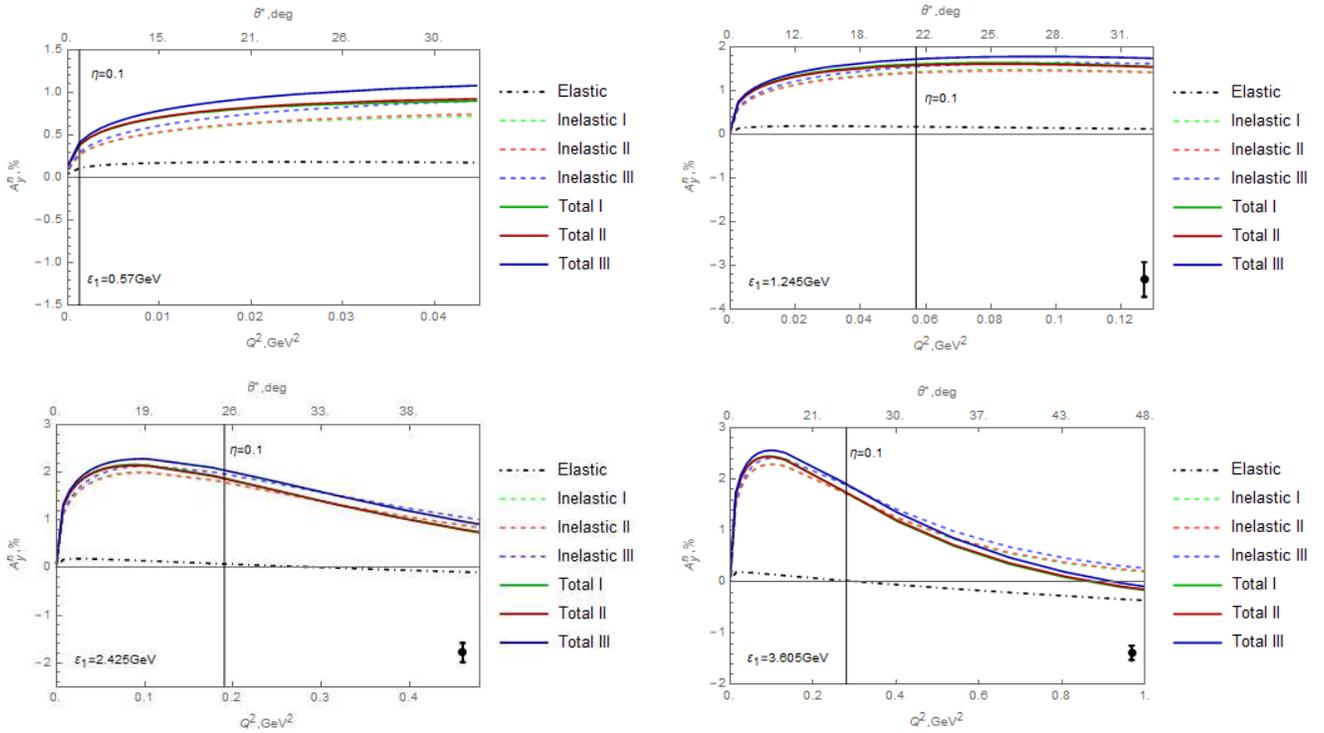}
    \caption{\label{fig:3} Neutron-normal single-spin asymmetry predictions. Vertical lines represent the $\eta = 0.1$ constraint. The datapoints are from Ref. \cite{ZhangSSA2015}.}
\end{figure*}

The expressions given in Eq. (\ref{1.512}), which we call near forward parametrizations, may now be employed to evaluate contractions of the leptonic and hadronic tensors. These contractions may be simplified if one takes into account gauge invariance of electromagnetic interactions, which implies that
\begin{equation}\label{1.901}
\begin{split}
     L^{\mu \alpha \beta} q_{\mu} & = L^{\mu \alpha \beta} q_{1 \beta} = L^{\mu \alpha \beta} q_{2 \alpha} = 0, \\
     H_{\mu \alpha \beta} q^{\mu} & = H_{\mu \alpha \beta} q^{\beta}_1 = H_{\mu \alpha \beta} q^{\alpha}_2 = 0.
\end{split}
\end{equation}
Once the contractions are performed, the near forward behavior of the asymmetry can be calculated by taking numerically the integral in Eq. (\ref{1.501}). Since all three parametrizations $\hat{T}_{\alpha \beta}^{(\RN{1})}, \hat{T}_{\alpha \beta}^{(\RN{2})},$ and $\hat{T}_{\alpha \beta}^{(\RN{3})}$ are normalized to provide the correct forward limit expression, the respective asymmetry predictions $A_{y}^{N (\RN{1})}, A_{y}^{N (\RN{2})},$ and $A_{y}^{N (\RN{3})}$ appear to be equivalent and no preference may be given to any of them. We will employ the differences in predictions, obtained with parametrizations \RN{1}, \RN{2}, and \RN{3}, to quantitatively set a limitation on our approach. In order to do so, let us define the following variables:
\begin{equation}\label{1.9031}
\begin{split}
     \eta_{12} & \equiv \frac{\big| A_{y}^{N (\RN{1})} - A_{y}^{N(\RN{2})} \big|}{\mathrm{min} \big( \big|A_{y}^{N (\RN{1})} \big|, \big|A_{y}^{N (\RN{2})} \big| \big)},
\end{split}
\end{equation}
\begin{equation}\label{1.9032}
\begin{split}
     \eta_{13} & \equiv \frac{\big| A_{y}^{N (\RN{1})} - A_{y}^{N(\RN{3})} \big|}{\mathrm{min} \big( \big|A_{y}^{N (\RN{1})} \big|, \big|A_{y}^{N (\RN{3})} \big| \big)},
\end{split}
\end{equation}
\begin{equation}\label{1.9033}
\begin{split}
     \eta_{23} & \equiv \frac{\big| A_{y}^{N (\RN{2})} - A_{y}^{N(\RN{3})} \big|}{\mathrm{min} \big( \big|A_{y}^{N (\RN{2})} \big|, \big|A_{y}^{N (\RN{3})} \big| \big)},
\end{split}
\end{equation}
\begin{equation}\label{1.9034}
\begin{split}
     \eta & \equiv \mathrm{max} \big( \eta_{12},\eta_{13},\eta_{23} \big).
\end{split}
\end{equation}
Based on these definitions, the parameter $\eta$ is a theoretical error band of our model calculation.

\begin{figure*}[htp]
    \includegraphics[scale=0.45]{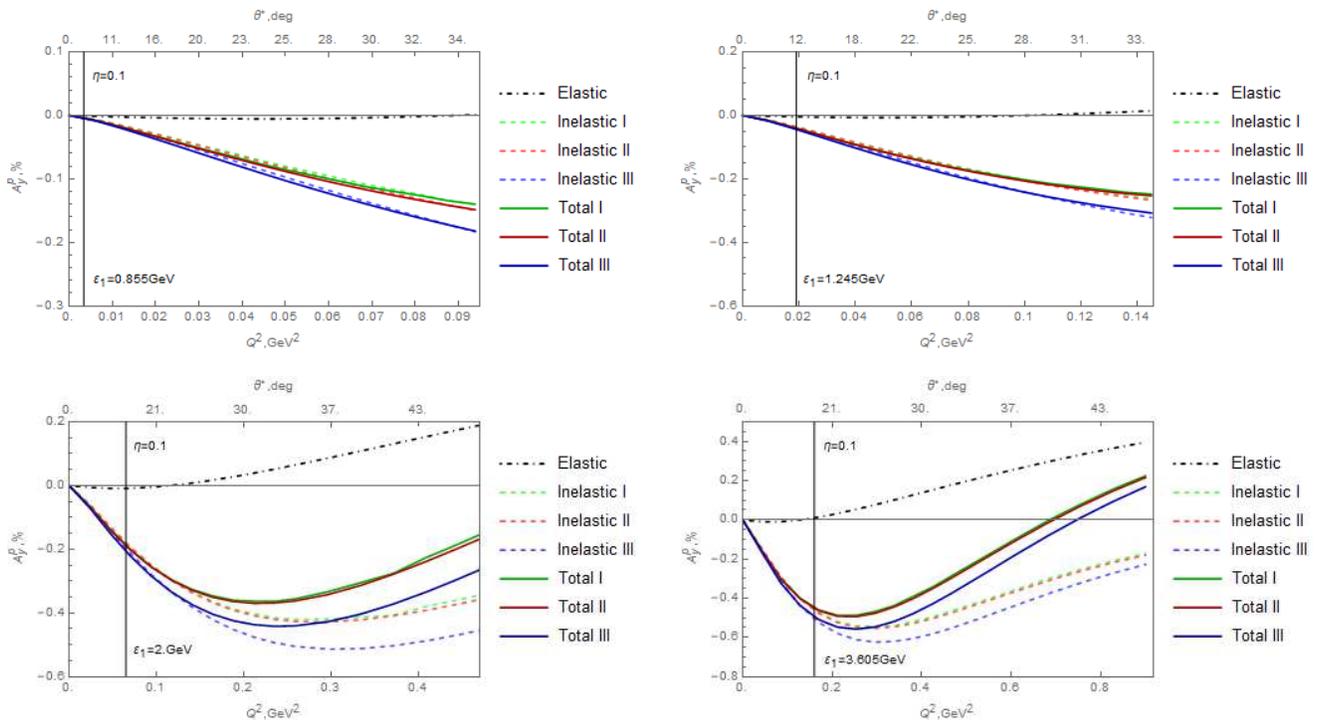}
    \caption{\label{fig:4} Proton-normal single-spin asymmetry predictions.}
\end{figure*}

In Figs. \ref{fig:3} and \ref{fig:4}, we display the target-normal SSA predictions for elastic $e^- n$ and $e^- p$ scattering, correspondingly. Vertical $\eta = 0.1$ lines on our plots indicate regions of kinematics (on the left side of these lines) for which $\eta$ is within 10\%. These are the regions of a desired theoretical uncertainty. Beyond the $\eta = 0.1$ line, the uncertainty of the near forward calculation becomes significant, indicating that we cannot rely on the usage of near forward contributions alone any longer. Unfortunately, as we can see from Fig. \ref{fig:3}, the constraint $\eta \leq 0.1$ implies that our results cannot be directly compared to the results of the experimental measurement of Ref. \cite{ZhangSSA2015} (we extend our theoretical curves beyond the $\eta = 0.1$ region just for illustrative purposes). More interestingly, we can also notice that the FL slope of theoretical curves is of an opposite sign as compared to that expected from the experiment. The only reasonable explanation for such a ``mismatch'' between existing experimental data and theoretical predictions of our model is that the neutron-normal measurement of Ref. \cite{ZhangSSA2015} was conducted in the kinematical region where additional, nonforward amplitudes take over. This also suggests that if someone were to perform a similar measurement - with an invariant $s$ being fixed at one of the values employed in Ref. \cite{ZhangSSA2015} and with an opportunity to access smaller values of $Q^2$ - they would find that the asymmetry $A_y^n$ crosses zero at least once in the interval $0 < Q^2 < 1$ GeV${}^2$. At sufficiently low $Q^2$, where the forward amplitudes eventually become dominant, the behavior of the asymmetry would be described by our model.

The proton-normal SSA predictions $A_y^p$, which are shown in Fig. \ref{fig:4}, are calculated analogously, one just needs to replace neutron form factors and scaling functions by corresponding proton quantities. Initial beam energies in Figs. \ref{fig:3} and \ref{fig:4} were chosen to perform a comparison of our calculations with experimental data of Ref. \cite{ZhangSSA2015}, with theoretical estimations of Ref. \cite{PasquiniSSA2004}, as well as to provide theory predictions for possible future measurements. As it can clearly be seen from our results, the FL slopes of both neutron- and proton-normal asymmetries calculated according to our approach are of the same sign as those of Ref. \cite{PasquiniSSA2004}. In addition, in the kinematical region of an overlap of both calculations, our predictions are of the same order of magnitude as those of Ref. \cite{PasquiniSSA2004}. The relative difference between theoretical curves is expected to be coming from the fact that the calculation of the letter reference is based on the $\gamma^* N \rightarrow \pi N$ electroabsorption amplitudes, whereas our approach employs deep inelastic structure functions, which contain information on the states beyond the $\pi N$ production threshold in the TPE loop.

The present calculation differs from that of De Rujula \textit{et al.} \cite{RujulaSSA1971} in several substantive aspects. First, we provide the near forward parametrization for $T_{\alpha \beta}$, whereas the authors of Ref. \cite{RujulaSSA1971} use the forward parametrization. Specifically, we do not resort to the $q_1 = q_2$ limit in our model for tensors $(\tau_1)_{\alpha \beta}, (\tau_3)_{\alpha \beta}, (\tau_4)_{\alpha \beta}, (\tau_{19})_{\alpha \beta}$. Moreover, in our parametrization for $W_{1 (2)}$ we provide an additional exponential suppression factor, which is derived from diffractive (near forward) Compton scattering measurements. Second, we use the most recent \cite{Christy} parametrizations for $F_{1}^{\mathrm{DIS}}$ and $F_{2}^{\mathrm{DIS}}$, whereas the authors of Ref. \cite{RujulaSSA1971} used parametrizations from early 70s. Finally, we introduced a quantitative criterium, $\eta$, which establishes an upper $Q^2$ bound on our model calculations. This bound prevents us from comparing our results vs. those presented in Table 1 of Ref. \cite{RujulaSSA1971}, as 7 (out of 9) datapoints given there fall into a region $\eta \gg 0.1$.

Finally, we would like to point out that the asymmetry $A_y^N$ is a function of two variables: the beam energy $\varepsilon_1$ and the scattering angle $\theta$. Since both variables are frame-dependent quantities, it is more illustrative to consider the asymmetry as a function of two invariants: $s$ and $Q^2$, for example. This means that there can be plotted a distinct $Q^2$ dependence of $A_y^N$ for every chosen value of $s$. For this reason, we provide four different proton- and neutron-normal asymmetry predictions, which correspond to four different values of $\varepsilon_1$.

\section{Conclusions}\label{1.1000}

We have computed the target-normal single-spin asymmetry at near forward angles in elastic electron-proton and electron-neutron scattering. Neglecting higher-order effects, this asymmetry is provided by the interference between one- and two-photon exchange amplitudes. The TPE amplitude includes both elastic and inelastic loop contributions, and the calculation of the latter one requires model assumptions. We have constructed the respective inelastic VVCS tensor that can be used in the kinematical region characterized by the $Q^2 \ll s$ condition. Our parametrization takes into account the following considerations: (a) exact knowledge of tensor structures contributing to the general nonforward VVCS tensor, (b) exact knowledge of the relationship between the imaginary part of the forward scattering amplitude and respective structure functions measurable in deep inelastic scattering experiments. To describe the $Q^2$ dependence of the asymmetry near its forward limit ($Q^2 = 0$), we made use of the experimentally known slope of the Compton scattering differential cross section. Our unitarity-based calculation features the following properties: it is properly normalized to provide a correct forward limit expression for the VVCS tensor, it is symmetric with respect to exchanges of virtual photons in the two-photon loop, and it improves on previous high energy ($Q^2 \ll s$) parametrizations of the VVCS tensor by directly accounting for the contributions coming from longitudinal photon exchanges in the TPE loop. Moreover, besides simply mentioning $Q^2 \ll s$ as a kinematical constraint on our approach, we have also suggested a quantitative criterium that can be used for the determination of the upper $Q^2$ bound in predictions of $A_y^N$ for any chosen value of the initial beam energy.

We found that in the kinematical range of an overlap of our calculations and calculations of Ref. \cite{PasquiniSSA2004}, the target-normal asymmetry predictions of both approaches appear to be in a reasonable agreement with each other. It would also be interesting to compare our results with the near-forward approximation model of Ref. \cite{TomalakVVCS2016}, provided that it is applied to predict the target-normal SSA. The comparison between experimental data of Ref. \cite{ZhangSSA2015} and our predictions, which are based on an extrapolation of near forward asymmetries from their forward limit expressions, shows a disagreement between the experiment and our theory. In contrast, the beam-normal SSA predictions that are based on a similar theoretical extrapolation procedure \cite{AfanasevBNSSA2004, GorchteinDispersive2007, GorsteinBNSSA2008} are in a good agreement with experiments \cite{G0BNSSA2007, HAPPEX2012}. Therefore, in order to compare experimental measurements with our theory in kinematics of the JLab experiment \cite{ZhangSSA2015}, it is required to completely model the nonforward Compton amplitude in our theoretical estimations. Alternatively, it would be desirable to extend future experimental measurements of the target-normal SSA to smaller scattering angles, at which the direct comparison between the experiment and the current theory is possible.

\acknowledgments
We are grateful to M.E. Christy, I. Lavrukhin and M. Mai, for useful discussions. We thank O. Tomalak for a careful review of our preprint and pointing out a typo in units of constant $B$. The Feynman diagrams in this paper were prepared using JaxoDraw \cite{JaxoDraw}. This work was supported in part by The George Washington University through the Gus Weiss endowment, in part by a JSA/JLab Graduate Fellowship Award, and in part by the National Science Foundation under Grant No. PHY-1812343.
\appendix

\section{Center-of-mass frame notations and relations}\label{1.600}

Using the notation introduced in Sec. \ref{1.100}, c.m. components of respective four-vectors can be defined as
\begin{equation}\label{1.601}
\begin{split}
    k_1 & = (\epsilon_1^*, \vec{k}_1^*), \ \ \ \ \ \  k_2 = (\epsilon_2^*, \vec{k}_2^*), \\
    p_1 & = (E_1^*, - \vec{k}_1^*), \ \ \ p_2 = (E_2^*, - \vec{k}_2^*), \\
    K & = (\xi^*, \vec{K}^*), \ \ \ \ \ P = (\Sigma^*, - \vec{K}^*).
\end{split}
\end{equation}

In the ultrarelativistic approximation ($\epsilon_1^* \gg m$), the inelastic scattering process
\begin{equation}\label{1.602}
    l(k_1) + N(p_1) \rightarrow l(K) + X(P)
\end{equation}
can be described by the following components of four-vectors given in the invariant form \cite{Bykling1973}:
\begin{equation}\label{1.603}
\begin{split}
    \epsilon_1^* & = |\vec{k}_1^*| = \frac{s - M^2}{2 \sqrt{s}}, \ \ E_1^* = \frac{s + M^2}{2 \sqrt{s}}, \\
    \xi^* & = |\vec{K}^*| = \frac{s - W^2}{2 \sqrt{s}}, \ \ \Sigma^* = \frac{s + W^2}{2 \sqrt{s}}.
\end{split}
\end{equation}
The elastic process
\begin{equation}\label{1.604}
    l(k_1) + N(p_1) \rightarrow l(k_2) + N(p_2)
\end{equation}
represents a special case ($X = N$, $W^2 = M^2$) of the inelastic process (\ref{1.602}). As a result, one finds that
\begin{equation}\label{1.605}
\begin{split}
    \epsilon_1^* & = |\vec{k}_1^*| \equiv \epsilon^* = \frac{s - M^2}{2 \sqrt{s}}, \\
    E_1^* & \equiv E^* = \frac{s + M^2}{2 \sqrt{s}}.
\end{split}
\end{equation}

If one chooses to perform the integration in Eq. (\ref{1.501}) in terms of c.m. frame variables, it is important to establish certain relations between corresponding integration parameters. In order to do so, we note that when the c.m. coordinate system is oriented as it is shown in Fig. \ref{fig:2}, and $\theta^*$ represents the respective c.m. scattering angle, we can write that
\begin{equation}\label{1.607}
\begin{split}
    k_1 & = (\epsilon^*, 0, 0, |\vec{k}^*|), \\
    k_2 &= (\epsilon^*, |\vec{k}^*| \sin \theta^*, 0, |\vec{k}^*| \cos \theta^*).
\end{split}
\end{equation}
In addition, we define $\phi_1$ to be the azimuthal angle of the intermediate electron state, and $\theta_1^* \equiv \angle \big( \vec{k}_1^*, \vec{K}^* \big)$ and $\theta_2^* \equiv \angle \big( \vec{k}_2^*, \vec{K}^* \big)$ to be its polar angles. With these definitions, the four-momentum of the intermediate electron can be written as
\begin{equation}\label{1.608}
    K = (\xi^*, |\vec{K}^*| \sin \theta_1^* \cos \phi_1^*, |\vec{K}^*| \sin \theta_1^* \sin \phi_1^*, |\vec{K}^*| \cos \theta_1^*).
\end{equation}
Moreover, using the identity $\vec{K}^* \cdot \vec{k}_2^* = |\vec{K}^*| |\vec{k}_2^*| \cos \theta_2^*$, one may find that
\begin{equation}\label{1.609}
    \cos \theta_2^* = \cos \theta^* \cos \theta_1^* + \sin \theta^* \sin \theta_1^* \cos \phi_1^*.
\end{equation}
The momentum transfer $Q^2$ and virtualities $Q_1^2, Q_2^2$ defined in Eq. (\ref{1.103}), are then given by
\begin{equation}\label{1.606}
\begin{split}
    Q^2 & = \frac{1}{2 s} (s - M^2)^2 (1 - \cos \theta^*), \\
    Q_1^2 & = \frac{1}{2 s} (s - M^2) (s - W^2) (1 - \cos \theta_1^*), \\
    Q_2^2 & = \frac{1}{2 s} (s - M^2) (s - W^2) (1 - \cos \theta_2^*).
\end{split}
\end{equation}

\section{High-energy electron-nucleon scattering}\label{1.700}

\begin{figure}[htp]
    \includegraphics[scale=0.5]{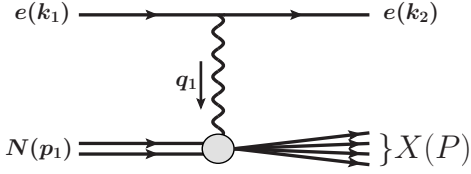}
    \caption{\label{fig:5} Deep inelastic scattering process}
\end{figure}

High-energy electron-nucleon scattering (deep inelastic scattering) plays a key role in determining the structure of the nucleon.
In the one-photon approximation, the lab frame double-differential cross section for the deep inelastic electron-nucleon scattering process, depicted in Fig. \ref{fig:5}, can be written as
\begin{equation}\label{1.701}
   \bigg (\frac{d \sigma_{1 \gamma}}{d \Omega d\varepsilon_2} \bigg)^{lab} = \frac{\alpha^2}{q_1^4} \frac{\varepsilon_2}{\varepsilon_1} \frac{1}{4 \pi M} \ l^{\alpha \beta} W_{\alpha \beta}^{\mathrm{DIS}},
\end{equation}
where $\varepsilon_1$ ($\varepsilon_2$) is the lab frame energy of the initial (final) electron. In addition, if we are considering the scattering of the unpolarized electron off the polarized nucleon target and the polarizations of the final particles are not measured, then the leptonic $l^{\alpha \beta}$ and the DIS hadronic $W_{\alpha \beta}^{\mathrm{DIS}}$ tensors are given by
\begin{multline}\label{1.702}
     l^{\alpha \beta} \equiv \frac{1}{2} \sum \limits_{S_e, S'_e} \bar{u}(k_1, S_e) \gamma^\alpha u(k_2, S'_e) \bar{u}(k_2, S'_e) \gamma^\beta u(k_1, S_e) \\
                       = \frac{1}{2} \mathrm{Tr} \Big[ (\slashed{k}_1 + m) \gamma^\alpha (\slashed{k}_2 + m) \gamma^\beta \Big] \\
                       = 2 \Big[ k_1^\alpha k_2^\beta + k_2^\alpha k_1^\beta - (k_1 \cdot k_2) g^{\alpha \beta} + m^2 g^{\alpha \beta} \Big],
\end{multline}
\begin{equation}\label{1.703}
\begin{split}
     W_{\alpha \beta}^{\mathrm{DIS}} & \equiv \sum \limits_X (2 \pi)^4 \delta^{(4)}\Big( p_1 + q_1 - P \Big) \\
     & \cdot <p_1, S_N|J^{\dag}_\alpha (0)|X> <X|J_{\beta} (0)|p_1, S_N>.
\end{split}
\end{equation}
It appears that the DIS hadronic tensor can be split into a sum of a symmetric and antisymmetric parts \cite{Ioffe1984}
\begin{equation}\label{1.704}
\begin{split}
     W_{\alpha \beta}^{\mathrm{DIS}} & \equiv W_{\alpha \beta}^{\mathrm{S}}  + i \ W_{\alpha \beta}^{\mathrm{A}}.
\end{split}
\end{equation}
The symmetric part is independent of the nucleon's spin, whereas polarization effects are described by the antisymmetric part. With our definitions of the leptonic and DIS hadronic tensors as in Eqs. (\ref{1.702}) and (\ref{1.703}), respectively, it is common to use the following parametrization for the symmetric and antisymmetric parts of the DIS hadronic tensor:
\begin{equation}\label{1.706}
\begin{split}
     & \frac{1}{4 \pi M} W_{\alpha \beta}^{\mathrm{S}} \equiv \Big( - g_{\alpha \beta} + \frac{q_{1 \alpha} q_{1\beta}}{q_1^2} \Big) W_1^{\mathrm{DIS}} (q_1^2, \nu_1) \\
     & \hspace{0.25in} + \frac{1}{M^2}  \Big(p_{1 \alpha} - \frac{(p_1 \cdot q_1)}{q_1^2} q_{1 \alpha} \Big) \Big( p_{1 \beta} - \frac{(p_1 \cdot q_1)}{q_1^2} q_{1 \beta} \Big)\\
     & \hspace{0.25in} \cdot  W_2^{\mathrm{DIS}} (q_1^2, \nu_1),
\end{split}
\end{equation}
\begin{equation}\label{1.707}
\begin{split}
     & \frac{1}{4 \pi M} W_{\alpha \beta}^{\mathrm{A}} \equiv \varepsilon_{\alpha \beta \rho \sigma } q_1^\rho \Big[ M S_N^\sigma  G_1^{\mathrm{DIS}} (q_1^2, \nu_1) \\
     & \hspace{0.75in} + \nu_1 \Big(S_N^\sigma - \frac{(S_N \cdot q_1)}{(p_1 \cdot q_1)} p_1^\sigma \Big) G_2^{\mathrm{DIS}} (q_1^2, \nu_1) \Big],
\end{split}
\end{equation}
where $W_{1,2}^{\mathrm{DIS}}$ and $G_{1,2}^{\mathrm{DIS}}$ are the nucleon unpolarized and polarized (or spin) structure functions, respectively, and $\nu_1 = (p_1 \cdot q_1) / M$.

The nucleon unpolarized structure functions, which we use in our calculations, are related to the absorption cross sections $\sigma_T$ and $\sigma_L$ of virtual transverse and longitudinal photons, respectively, via
\begin{equation}\label{1.1105}
\begin{split}
    & W_1^{\mathrm{DIS}}(q_1^2, \nu_1) = \frac{\nu_1}{4 \pi^2 \alpha} \sigma_T (q_1^2, \nu_1), \\
    & W_2^{\mathrm{DIS}}(q_1^2, \nu_1) = \frac{q_1^2 \nu_1 \Big( \sigma_T (q_1^2, \nu_1) + \sigma_L (q_1^2, \nu_1) \Big)}{4 \pi^2 \alpha (q_1^2 - \nu_1^2)} .
\end{split}
\end{equation}
Sometimes, it is convenient to relate the above-mentioned structure functions to corresponding dimensionless scaling functions $F_{1,2}^{\mathrm{DIS}}$ by introducing the Bjorken scaling variable $x_B$ instead of variable $\nu$
\begin{equation}\label{1.709}
     x_{B_1} \equiv - \frac{q_1^2}{2 M \nu_1}.
\end{equation}
The scaling functions are defined by
\begin{equation}\label{1.710}
\begin{split}
     M W_1^{\mathrm{DIS}}(q_1^2, \nu_1) & \equiv  F_1^{\mathrm{DIS}}(q_1^2, x_{B_1}), \\
     \nu_1 W_2^{\mathrm{DIS}}(q_1^2, \nu_1) & \equiv F_2^{\mathrm{DIS}}(q_1^2, x_{B_1}).
\end{split}
\end{equation}

\section{Forward VVCS amplitude}\label{1.1100}
\begin{figure}[htp]
    \includegraphics[scale=0.5]{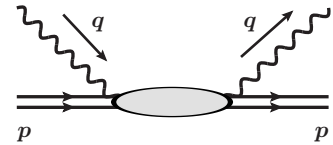}
    \caption{\label{fig:6} Forward Compton scattering kinematics.}
\end{figure}
In the forward kinematics ($t=0$), using the notation shown in Fig. \ref{fig:6}, the VVCS amplitude looks like
\begin{equation}\label{1.1101}
\begin{split}
    & T_{\alpha \beta} \Big|_{\mathrm{FL}} = \Big(- q^2 g_{\alpha \beta} + q_{\alpha} q_{\beta} \Big) T_1 (q^2, \nu) \\
    & \hspace{0.25in} + \frac{1}{M^2} \Big( q^2 p_\alpha - (p \cdot q) q_\alpha \Big) \Big(q^2 p_\beta - (p \cdot q) q_\beta \Big) T_2 (q^2, \nu) \\
    & \hspace{0.55in} + \frac{i}{M} \gamma_{\alpha \beta \rho} q^\rho S_1 (q^2, \nu) \\
    & \hspace{0.25in} + \frac{i}{2 M^2} \Big( q^2 \gamma_{\alpha \beta} + q_\alpha \gamma_{\beta \rho} q^\rho - q_\beta \gamma_{\alpha \rho} q^\rho \Big) S_2 (q^2, \nu),
\end{split}
\end{equation}
where we used the following definitions:
\begin{equation}\label{1.1102}
\begin{split}
    & \gamma_{\alpha \beta \rho} \equiv \epsilon_{\alpha \beta \rho \sigma} \gamma^\sigma \gamma^5, \\
    & \gamma_{\alpha \beta} \equiv  \epsilon_{\alpha \beta \rho \sigma} \gamma^\rho \gamma^\sigma \gamma^5, \\
    & \nu \equiv \frac{(p \cdot q)}{M} = \frac{W^2 - M^2 - q^2}{2 M}.
\end{split}
\end{equation}
Moreover, $T_{1,2}$ are the so-called spin-independent and $S_{1,2}$ the spin-dependent invariant amplitudes.

One can notice that the VVCS amplitude consists of the symmetric and antisymmetric parts, $T_{\alpha \beta} = T_{\alpha \beta}^S + i T_{\alpha \beta}^A$, where
\begin{equation}\label{1.1103}
\begin{split}
    & T_{\alpha \beta}^S \Big|_{\mathrm{FL}} = \Big( - q^2 g_{\alpha \beta} + q_{\alpha} q_{\beta} \Big) T_1 (q^2, \nu) \\
    & \hspace{0.25in} + \frac{1}{M^2} \Big( q^2 p_\alpha - (p \cdot q) q_\alpha \Big) \Big( q^2 p_\beta - (p \cdot q) q_\beta \Big) T_2 (q^2, \nu),
\end{split}
\end{equation}
\begin{equation}\label{1.11033}
\begin{split}
    & T_{\alpha \beta}^A \Big|_{\mathrm{FL}} = \frac{1}{M} \gamma_{\alpha \beta \rho} q^\rho S_1 (q^2, \nu) \\
    & \hspace{0.25in} + \frac{1}{2 M^2} \Big( q^2 \gamma_{\alpha \beta} + q_\alpha \gamma_{\beta \rho} q^\rho - q_\beta \gamma_{\alpha \rho} q^\rho \Big) S_2 (q^2, \nu).
\end{split}
\end{equation}
The optical theorem relates the imaginary parts of the forward amplitudes $T_{1,2}$ to the nucleon unpolarized structure functions $W_{1, 2}^{\mathrm{DIS}}$
\begin{equation}\label{1.1104}
\begin{split}
    & \mathrm{Im} \Big[ T_1 (q^2, \nu) \Big] = \frac{\pi}{q^2} W_1^{\mathrm{DIS}} (q^2, \nu), \\
    & \mathrm{Im} \Big[ T_2 (q^2, \nu) \Big] = \frac{\pi}{q^4} W_2^{\mathrm{DIS}} (q^2, \nu).
\end{split}
\end{equation}

\begin{widetext}

\section{Tensor structures}\label{1.800}

Here we show the tensor structures:
\begin{equation}\label{1.801}
    (\tau_1)_{\alpha \beta} = - (q_1 \cdot q_2) g_{\alpha \beta} + q_{1 \alpha} q_{2 \beta},
\end{equation}
\begin{equation}\label{1.802}
    (\tau_3)_{\alpha \beta} =  q_1^2 q_2^2 g_{\alpha \beta} + (q_1 \cdot q_2) q_{2 \alpha} q_{1 \beta} - \frac{q_1^2 + q_2^2}{2} (q_{1 \alpha} q_{1 \beta} + q_{2 \alpha} q_{2 \beta}) + \frac{q_1^2 - q_2^2}{2} (q_{1 \alpha} q_{1 \beta} - q_{2 \alpha} q_{2 \beta}),
\end{equation}
\begin{equation}\label{1.803}
\begin{split}
    (\tau_4)_{\alpha \beta} = (\bar{p} \cdot \bar{q}) (q_1^2 + q_2^2) g_{\alpha \beta} - (\bar{p} \cdot \bar{q}) (q_{1 \alpha} q_{1 \beta} + q_{2 \alpha} q_{2 \beta}) - \frac{q_1^2 + q_2^2}{2} (q_{1 \alpha} \bar{p}_\beta + \bar{p}_\alpha q_{2 \beta}) + & \frac{q_1^2 - q_2^2}{2} (q_{1 \alpha} \bar{p}_\beta - \bar{p}_\alpha q_{2 \beta}) \\
    & + (q_1 \cdot q_2) (q_{2 \alpha} \bar{p}_\beta + \bar{p}_\alpha q_{1 \beta}),
\end{split}
\end{equation}
\begin{equation}\label{1.804}
     (\tau_{19})_{\alpha \beta} = 2 (\bar{p} \cdot \bar{q})^2 q_{2 \alpha} q_{1 \beta} + 2 q_1^2 q_2^2 \bar{p}_\alpha \bar{p}_\beta - (\bar{p} \cdot \bar{q}) (q_1^2 + q_2^2)(q_{2 \alpha} \bar{p}_\beta + \bar{p}_\alpha q_{1 \beta}) - (\bar{p} \cdot \bar{q}) (q_1^2 - q_2^2)(q_{2 \alpha} \bar{p}_\beta - \bar{p}_\alpha q_{1 \beta}).
\end{equation}
\end{widetext}

\bibliography{NormalSSA}

\end{document}